\documentclass[letterpaper,sort&compress]{aipproc}
\layoutstyle{8x11single}
\def\gtrsim{\mathrel{\hbox{\rlap{\hbox{\lower4pt\hbox{$\sim$}}}\hbox{$>$}}}}
\def\lesssim{\mathrel{\hbox{\rlap{\hbox{\lower4pt\hbox{$\sim$}}}\hbox{$<$}}}}

\begin{document}
\title[BH-NS Merger Calculations]{Black Hole-Neutron Star Binary Merger Calculations: GRB Progenitors and the Stability of Mass Transfer}

\author{Joshua A.~Faber}{
  address={Dept. of Physics, University of Illinois at
  Urbana-Champaign, Urbana, IL 61801, USA},altaddress={NSF Astronomy
  and Astrophysics Postdoctoral Fellow},email={jfaber@uiuc.edu}
}
\author{Thomas W.~Baumgarte}{
address={Dept. of Physics and Astronomy, Bowdoin College, Brunswick,
  ME 04011 USA}}
\author{Stuart L.~Shapiro}{
address={Dept. of Physics, University of Illinois at
  Urbana-Champaign, Urbana, IL 61801, USA}, altaddress={Dept. of
  Astronomy and NCSA, UIUC}}
\author{Keisuke Taniguchi}{
address={Dept. of Physics, University of Illinois at
  Urbana-Champaign, Urbana, IL 61801, USA}}
\author{Frederic A.~Rasio}{
address={Dept. of Physics and Astronomy, Northwestern University,
  Evanston, IL 60208 USA}}
\begin{abstract}
We have calculated the first dynamical evolutions of merging black
hole-neutron star binaries that treat the combined spacetime in a
nonperturbative general relativistic framework. Using the conformal flatness
approximation, we have studied how the location of the tidal
disruption radius with respect to the the black hole horizon and
innermost stable circular orbit (ISCO) affects the qualitative evolution of
the system. Based on simple arguments, we show that for a binary mass
ratio $q\gtrsim 0.24$, tidal disruption occurs outside the ISCO, while the
opposite is true for $q\lesssim 0.24$. When tidal disruption occurs
sufficiently far outside the ISCO, mass is transferred unstably from
the neutron 
star to the black hole, resulting in the complete disruption of the
neutron star.  When tidal disruption occurs slightly within the ISCO,
we find that some of the mass forms an extremely hot disk around the
black hole.  The resulting configurations in this case
are excellent candidates for
the progenitors of short-hard gamma ray bursts.
\end{abstract}
\date{\today}
\maketitle

\section{Introduction}
The mergers of compact object binaries consisting of either neutron
stars (NSs) or
black holes (BHs) are promising sources of
gravitational radiation, which hopefully will be detected soon by both
ground-based interferometers such as LIGO \cite{LIGO} and 
the proposed space-based mission LISA
\cite{LISA}. Understanding the evolution of these systems during and
after merger, including their emission in gravitational and, for
systems containing NSs
electromagnetic and neutrino radiation, requires a careful treatment of the
relativistic dynamics (see \cite{BS} for a recent review).  
A great deal of effort over the past decade has gone into studying NSNS
mergers, and recent treatments incorporate both fully general relativistic
gravitation as well as physically realistic nuclear equations
of state (EOS) \cite{STU,ST}.  More recently, there has been a
rapid burst of progress on numerical evolution of merging
BHBH binaries.  \citet{Pretorius} has
performed a stable evolution of the merger of two equal-mass
BHs, through the implementation of generalized harmonic coordinates.
Several groups have reported comparable results for binary
``puncture'' BHs, reporting numerical implementations based on the
BSSN formulation of Einstein's equations \cite{SN0,BS0}, which seem to
be considerably more stable than any previous treatment
\cite{God1,God2,UTB1,UTB2,PSU}. 

By contrast, BHNS mergers have received significantly less attention,
even though they represent a significant fraction of compact object
mergers visible in gravity waves (GWs) \cite{BKB}.  With the
exception of the work described herein \citep[hereafter papers 1 and
  2, respectively]{FBSTR,FBST}, all other calculations of BHNS 
mergers involve either Newtonian or pseudo-Newtonian gravitational
treatments of the NS self-gravity \cite{PSSPH}, 
or the entire binary \cite{JERF,LK,Ross}.
However, as we show here, general relativistic effects play an
extremely important role during the tidal disruption of the NS, which
occurs in the strong-field regime.  This is especially true when the
NS disrupts within the ISCO, since the deep relativistic gravitational
potential decreases the mass of ejected material while simultaneously
increasing the characteristic energy scale associated with it. 
In order to perform these calculations, we use fully relativistic
initial data that solve
the conformal thin-sandwich (CTS) equations for quasi-equilibrium BHNS
configurations in circular orbits, under the assumption that the BH mass is
significantly larger than that of the NS ($q\equiv M_{\rm NS}/M_{\rm
  BH}\ll 1$).  These initial data were first derived for synchronized
configurations in \cite{BSS}, and then extended to irrotational
configurations in \cite{TBFS}. 

A proper study of BHNS mergers is even more important today in light
of the recent localizations of short-hard gamma-ray bursts (SGRBs) for
the first time.  Long-soft bursts had been previously localized, and
several had been seen in coincidence with Type Ib/c supernovae,
indicating that they represented the collapse of massive stars
\cite{Paczynski98}. By observing the X-ray afterglow of SGRBs, the
     {\it Swift} and {\it 
  HETE-2} satellites have finally allowed us to identify where SGRBs
occur, and in many cases indicate the galaxy that served as host.  The
rapidly growing list of localized bursts now includes 
GRBs 050509b, 050724, 050813, and 051221 seen by {\it 
Swift}, and GRB 050709 seen by {\it HETE-2}.  Details about each of these, including
the physical parameters observed and inferred from these bursts, as
  well as a complete set of observational references, can be 
found in \citet{Berger3}.  In all cases, the inferred host had a
rather low star-formation rate: nearly zero for the putative hosts of
  050509b, 050724, and 050813, while significantly higher but still
  low relative to long GRB hosts for 050709 and 051221
  \citep{Hjorth,Soderberg}.   
Massive stars have very short lifetimes and are typically found in
  regions with high star formation, so  these observations favor the
  identification of a compact object 
binary merger as the progenitor, as originally suggested by
  \citet{Pac86}. A significant fraction of both neutron star-neutron
star (NSNS) and black hole-neutron star (BHNS) binaries, on the other hand, will take longer
than 1 Gyr between formation and merger \citep{BBR}.  

While our work here has focused on BHNS mergers, we note that fully
general relativistic calculations of NSNS mergers suggest that they
are also strong candidates for producing SGRBs.  The standard
schematic model for merger-induced short-duration GRBs 
involves creating a hot, massive ($M>0.01 M_{\odot}$) 
accretion disk around a
spinning BH (see \citet{Piran} for a thorough review).  Recent
fully general relativistic calculations show that such configurations
naturally result from either of two distinct formation channels in
NSNS mergers.  For NSs of significantly different masses (mass ratios
$q\lesssim 0.7$), tidal disruption of the lighter NS produces a heavy
disk that will accrete onto a newly formed BH at the binary
center-of-mass \cite{STU,ST}.  Alternately, for more equal-mass
systems, the merger may produce a hypermassive neutron star (HMNS),
supported against gravitational collapse by differential rotation with
a total mass heavier than the maximum allowed mass for uniformly
rotating configurations \cite{BSS00,Shap00}.  Magnetohydrodynamic
(MHD) simulations in
full GR show that the HMNS undergoes a delayed collapse, resulting in
a hot, magnetized torus surrounding a rotating BH, together with a
magnetic field collimated along the polar axis.  These conditions are
favorable for a burst powered either by neutrino annihilation or
MHD effects \citep{Duez,ShiUIUC}. 

As we discuss at length in Paper 2, it is very difficult to determine
whether current observations favor the likely progenitors of SGRBs to
be NSNS mergers, BHNS mergers, or possibly both.  While the most
recent calculations of the SGRB merger rate, ${\cal R}_{GRB}\sim 1-3~{\rm Myr^{-1}}$ per
Milky Way Galaxy \citep{Guetta2,Nakar} seem to be higher than
previously estimated, the predicted merger rates
from population synthesis calculations and other methods are still
uncertain by up to two orders of magnitude \cite{VT,BBR,KKL}.
Moreover, the projected distance from a localized SGRB to the center
of its host galaxy is insensitive to whether the merging binary is an
NSNS or BHNS 
system, since the projected distance likelihood functions for both
BHNS and NSNS mergers are very close for a wide range of galaxy masses
\citep{Bel06}. 

Perhaps the strongest constraint on possible GRB progenitors involves
the density of baryons around the central engine along the polar
axes. The low energy measured for GRB 050509b, $E_{\gamma}=3\times
10^{48}~{\rm erg}$, argues for an extremely 
low density of baryons surrounding the GRB, assuming that it occurred
at the inferred redshift $z=0.225$ and the Lorentz factor of the jet
$\Gamma$ is no bigger than the 
ratio $\eta$ of the energy in the jet to the rest energy of the
baryons through which it must travel, $\Gamma \le \eta\equiv
E_{\gamma}/M_{b}c^2$ \citep{ShemiPiran}.  For a typical Lorentz factor
$\Gamma\gtrsim 100$ \citep{Oech}, at most $10^{-8}
(\Omega/4\pi) M_{\odot}$ of material can surround the progenitor,
where $\Omega$ is the solid angle of the GRB jet.

\section{Numerical Techniques}

We performed dynamical 3+1-dimensional smoothed particle hydrodynamics (SPH)
calculations of BHNS mergers in the conformal flatness (CF)
approximation of GR \cite{Isen,WMM}.  For test masses in orbit about
Schwarzschild BHs our scheme identifies 
the relativistic ISCO exactly and accounts for relativistic dynamics
within the ISCO, unlike previous pseudo-Newtonian calculations
\citep{LK,JERF,Ross}.  Quasi-equilibrium initial data can also be
constructed to be conformally flat, so that they are consistent with
our evolution scheme.  During the dynamical evolution the CF
assumption is only approximate, since in general the spatial metric
would not remain conformally flat, but we expect these deviations to
be small. 
Our calculations of BHNS coalescence make use of a simplifying
technique introduced by \citet{BSS}:
by assuming an extreme mass ratio ($q\ll 1$), we may
hold the BH position fixed in space and solve the field equations for
the combined spacetime in a domain surrounding the NS but outside the
BH.  The resulting solution represents the fully dynamical
field configuration under our assumptions. 

Our numerical scheme is discussed in detail in Papers 1 and 2.  There, 
we evolved the matter adiabatically, but accounting for shocks,
by solving the coupled nonlinear elliptic
field equations with either the {\tt Lorene} spectral methods libraries,
publicly available at {\tt http://www.lorene.obspm.fr}, or via
iterated FFT convolution.  We include 
artificial viscosity
effects through a relativistic generalization of the form found in
\cite{HKAV}.

The defining parameter for determining the qualitative evolution of a
BHNS merger is the binary mass ratio $q\equiv M_{\rm NS}/M_{\rm BH}$.
Assuming that the tidal disruption of the NS begins at a separation
$a_{\rm R}$ where its volume in
isolation is equal to the volume of its Roche lobe,
\begin{equation}
a_{\rm R} / M_{\rm BH} = 2.17q^{2/3}(1+q)^{1/3}{\cal C}^{-1} \label 
{eq:ar}
\end{equation}
for a NS of compactness ${\cal C}\equiv M_{\rm NS}/R_{\rm NS}$ \citep 
{PacRoche},
using geometrized units with $G=c=1$.

Much of the previous discussion of BHNS mergers has portrayed the
outcome of these events as a simple process following one of two
evolutionary paths.  If disruption occurs outside the innermost stable
circular orbit (ISCO), the NS transfers mass onto the BH.  This widens
the orbit, leading to either a long-term stable mass transfer phase
\cite{CE,PZ}, or punctuated phases of Roche lobe overflow \cite{DLK}. 
For a sufficiently large BH (and thus small $q$), $a_{\rm R}$
is smaller than the innermost stable circular orbit (ISCO), $a_{\rm
ISCO}$, so that the NS passes through the ISCO before being tidally disrupted.
For a typical NS of compactness ${\cal C}=0.15$ the  
critical mass ratio at which tidal disruption occurs at $a_{\rm ISCO}$
is approximately $q_{crit}=0.24$.  For binaries with $q<q_{crit}$, it
has typically been 
suggested that the NS will be swallowed whole by the BH
\citep{Miller}. As we show below, these simple assumptions about
dynamics fail to describe the complicated processes seen in numerical calculations.

We cannot directly simulate the physical realistic ``critical case''
($q=q_c=0.24,~{\cal C}=0.15$) that disrupts at the ISCO, nor those
with lower-mass BHs in which the NS disrupts outside the ISCO, since they violate our
assumption of an extreme mass ratio.  We can, however, evolve
configurations with arbitrary values of $a_R$ 
by instead altering the NS compactness while holding the mass ratio
fixed at $q=0.1$.  In Paper 1, we studied the case where the NS
disrupts outside the ISCO by studying the evolution of a NS with
compactness $M/R=0.042$ in a binary of mass ratio $q=0.1$.  The NS was
taken to be initially corotating.  We consider first
the results from run B3a of Paper 1, which featured a NS
with a polytropic EOS with adiabatic index $\Gamma=2$. 
To study disruption within the ISCO we calculated the
evolution of binaries with the same mass ratio and NS EOS, but
compactness $0.14$ and $0.09$ (runs 1 and 2, respectively of Paper 2).
The latter of these is expected to disrupt at a binary separation
$a_R=5.3 M_{BH}$, just within the ISCO, and serves as our model for
the critical case.  For these two runs, the initial
configuration is irrotational, which is 
expected to be the physically realistic case. 

\section{Disruption outside the ISCO: Unstable mass transfer}\label{sec:outside}
\begin{figure}
\resizebox{.85\textwidth}{!}{\includegraphics{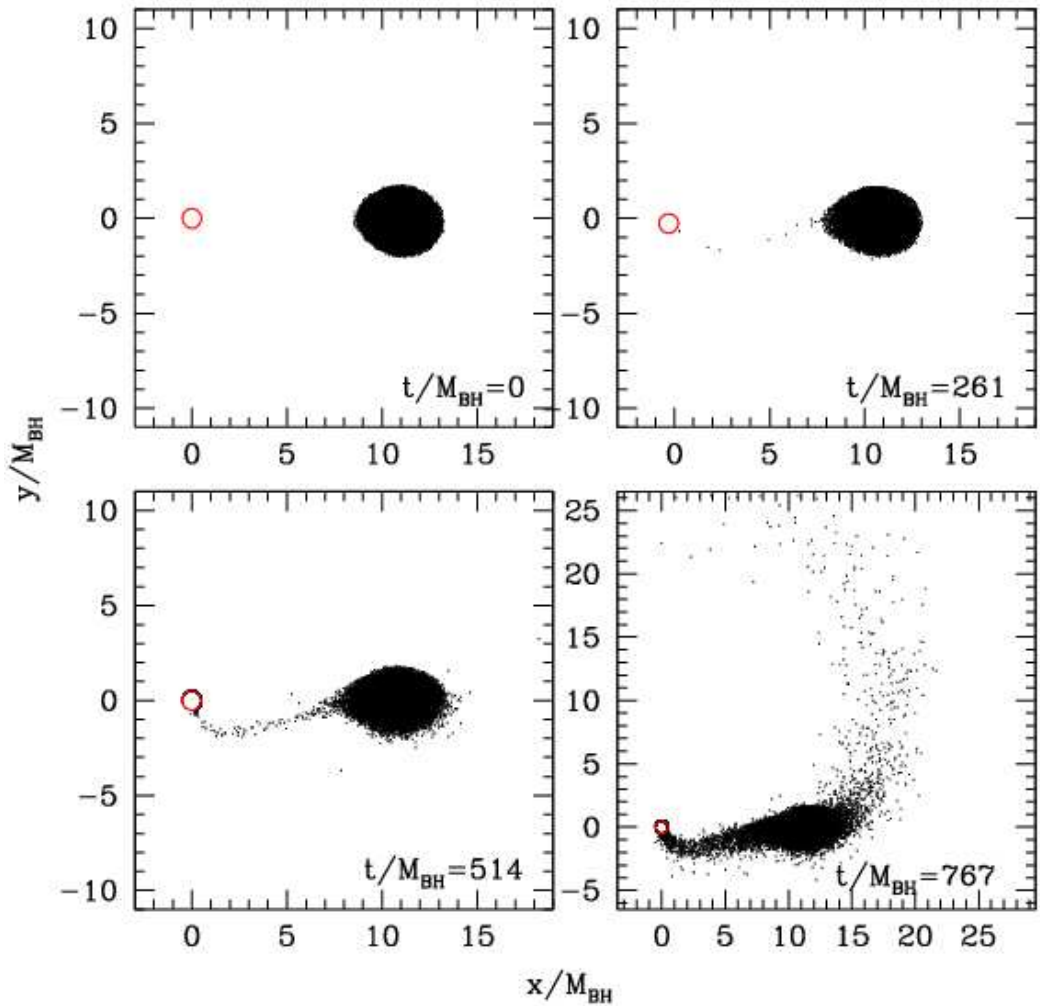}}
\caption{Snapshots of the fluid configuration projected into the
  orbital plane initially, and after 1, 2, and 3 full orbits. The NS
  has a compactness $M/R=0.042$ and the binary mass ratio 
  is $M_{\rm NS}/M_{\rm BH}=0.1$.   Mass lost through the inner
  Lagrange point accretes directly onto the BH for an orbit,
  at an increasing rate.  When the NS begins to expand, mass
  is lost through both the inner and outer Lagrange points, leading to
  the complete disruption of the NS.\label{fig:xybig}} 
\end{figure}

\begin{figure}
\resizebox{.85\textwidth}{!}{\includegraphics{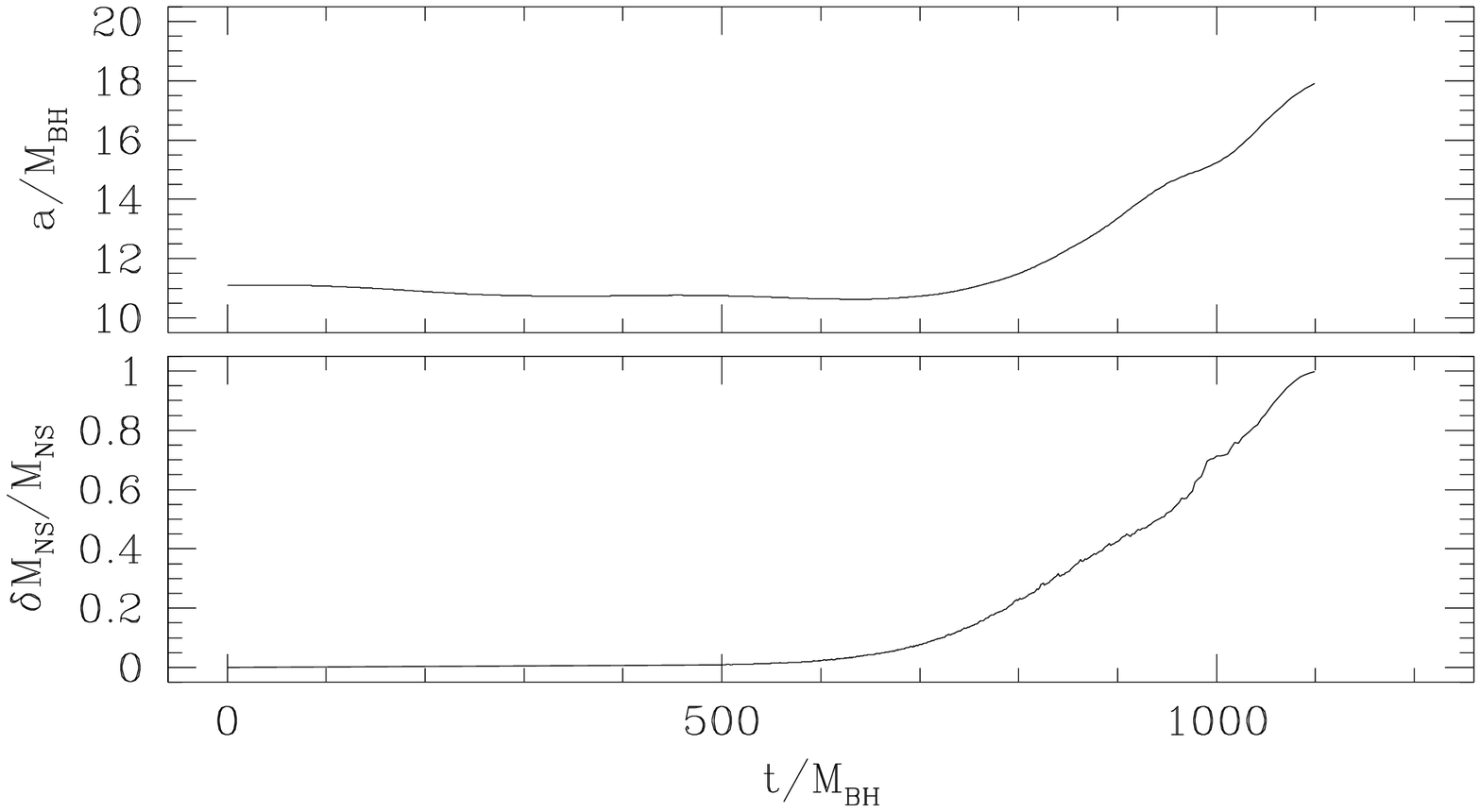}}
\caption{Binary separation $a/M_{\rm BH}$ (top) and fractional mass
  $\delta M/M_{\rm NS}$ lost from the NS (bottom) as a function of time, for
  the run 
  shown in Fig.~\protect\ref{fig:xybig}.  We see that mass loss 
is accompanied by a gradual increase in the orbital separation, but
  continues unquenched until the NS is completely disrupted.
\label{fig:mr}}
\end{figure}

For NS disruption outside the ISCO, the relevant timescales rule out
initially stable mass transfer.  When mass transfer begins, the NS has
a significant infall velocity, and it requires a significant fraction
of an orbit to reverse this velocity and allow the NS to spiral
outward.  During this time, mass loss through the inner Lagrange point
accelerates, stripping a great deal of matter from the NS.  Unless the
actual NS EOS is extremely stiff, mass transfer will not
stabilize, and the NS will be completely disrupted.  We illustrate
of this process in Figs.~\ref{fig:xybig} and \ref{fig:mr}. 

We see that mass transfer begins slowly, funneling matter through the
inner Lagrange point directly onto the BH.  During the first
orbit after this, the mass transfer rate increases, since the Lagrange
point actually moves inward into the outer layers of the NS.  Eventually, the NS
expands in response to mass loss, and matter is lost through the outer
Lagrange point as well.  The effect of the NS is catastrophic: matter
lost outward yields a net inward force that 
slows the widening of the orbit, preventing mass transfer from stabilizing.
As a result, the NS spirals outward at a rate too slow to quench the mass loss,
and is completely disrupted within another orbit.

We expect the evolution of the binary to follow this qualitative path
for BHs not much more massive than the NS ($q\gtrsim 0.3$), even if
the NS EOS is very stiff.  If mass loss can actually be quenched, it
will occur while the NS moves outward on an elliptical orbit.  Unlike the
model presented in \cite{DLK}, however, we expect that mass transfer
will begin again during the next approach toward pericenter, since
there is no obvious means to provide the angular momentum kick that
their model presupposes to boost the NS into a wider orbit. 

\section{Disruption inside the ISCO: Disk formation and Implications for SGRBs}
\label{sec:inside}

To study disruption within the ISCO, we consider two models, both
containing $n=1$ polytropes.  First is a NS of compactness $0.14$
in a binary of mass ratio  
$q=0.1$, for which $a_{\rm R}=3.2 M_{\rm BH}$ according to
Eq.~\ref{eq:ar}.  For this case, referred to as run 1 in Paper 2, the
NS is accreted completely by the BH, as tidal disruption
begins once the NS has already begun to approach the horizon.
Essentially no matter is ejected.  Such a model is unlikely
to produce a large luminosity in neutrinos, as the matter never shocks
as it falls onto the BH.  The GW signal from such an event may be
significant, both from the prior inspiral and the subsequent BH
ringing, but determining the exact GW
signal requires a fully non-conformally flat, 
general relativistic treatment that is beyond
the capabilities of our current code. 

\begin{figure}
\resizebox{.85\textwidth}{!}{\includegraphics{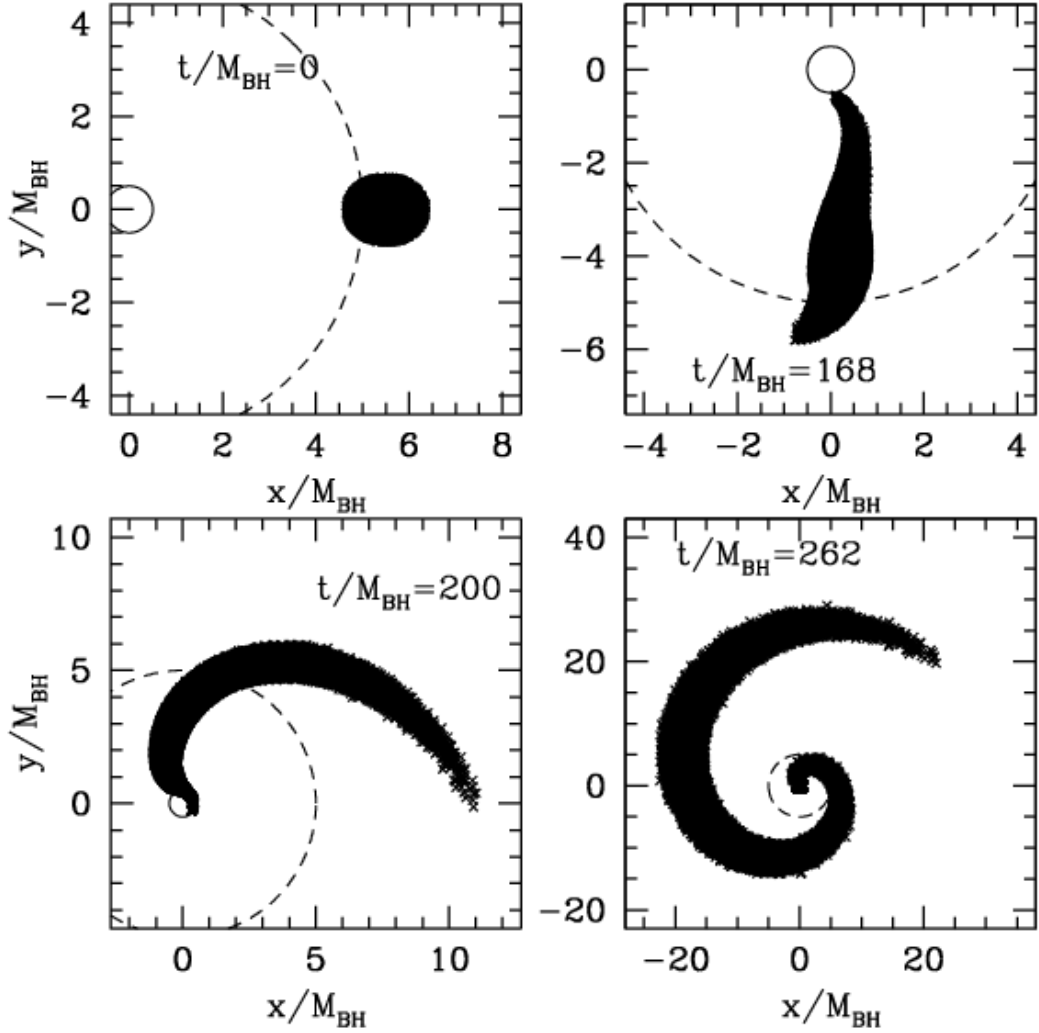}}
\caption{Snapshots of configurations taken from run 2 of Paper 2
  \protect\cite{FBST}, with conventions as in
  Fig.~\protect\ref{fig:xybig}.   
The NS has a compactness $M/R=0.09$ and the binary mass ratio
  is $M_{\rm NS}/M_{\rm BH}=0.1$.  We see the NS disrupts just within the
  ISCO (dashed curve), producing a mass-transfer stream outward, which
  eventually wraps around the BH (whose horizon is shown as the solid curve) to form a torus.  The
  initial orbital period is $P=105M_{\rm BH}$. \label{fig:xyz}}
\end{figure}

The more exciting case for producing an electromagnetic signal is run
2 of Paper 2, a NS of compactness $0.09$ in a binary of mass ratio
$q=0.1$, which disrupts just inside the ISCO, as shown in
Fig.~\ref{fig:xyz}. Note that radii in the figure are measured in
isotropic coordinates.  When tidal disruption begins after nearly two
orbits, as seen in the second panel, $98\%$ of the NS mass lies within
the ISCO.  From this point onward, however, rapid redistribution of
angular momentum allows a great deal more mass, $\sim 0.25 M_{\rm
  NS}$, to be expelled outside the ISCO, while the remaining fraction,
$0.75M_{\rm NS}$, is accreted promptly by the BH.

\begin{figure}
\resizebox{.85\textwidth}{!}{\includegraphics{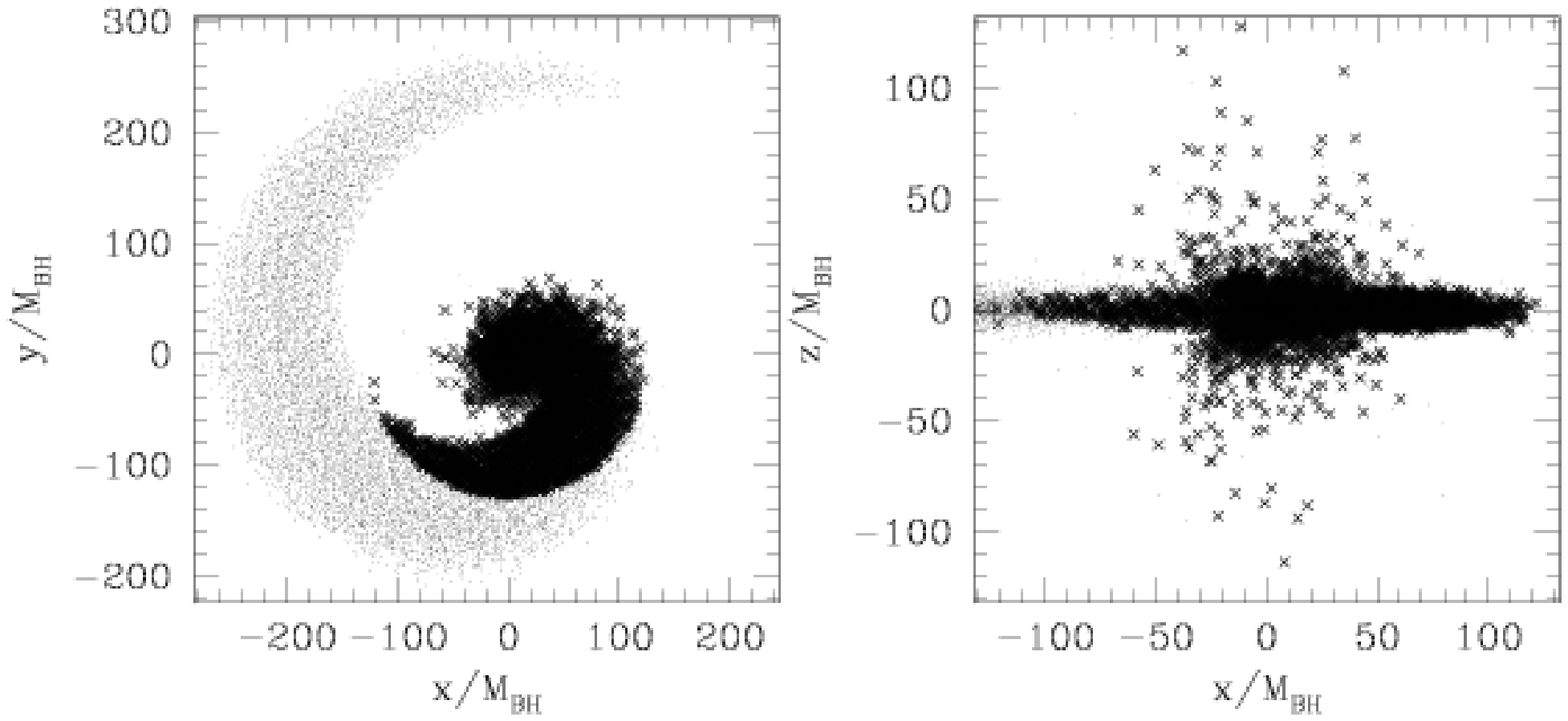}}
\caption{Matter configuration at the end of the simulation,
  $T=990M_{\rm BH}$, projected onto the equatorial (left panel) and
  meridional (right panel) showing the hot torus located within
  $r<50M_{\rm BH}$.  Bound fluid elements (satisfying $u_0-1<0$) are
  shown as crosses, unbound elements as points.  Note the different
  scales.\label{fig:xyz2}}
\end{figure}

While half of the ejected matter is unbound from the system, the
remaining fraction, $\sim 0.12 M_{\rm NS}$, will eventually fall back
and form a disk located just outside the ISCO.
This configuration satisfies all the geometric constraints
required for a GRB progenitor, as all matter lies in the equatorial
plane rather than the polar axis.
The bound matter in the disk is relatively cold at first, as the
matter in the arm is initially ejected without
strong shock heating.  Over time,
this disk generates heat via shocks and expands vertically into a torus
out to a radius of $r\sim 50M_{\rm BH}$ within $t=1000M_{BH}\sim
  0.07~{\rm s}$; (see Fig.~\ref{fig:xyz2}).

\begin{figure}
\resizebox{.85\textwidth}{!}{\includegraphics{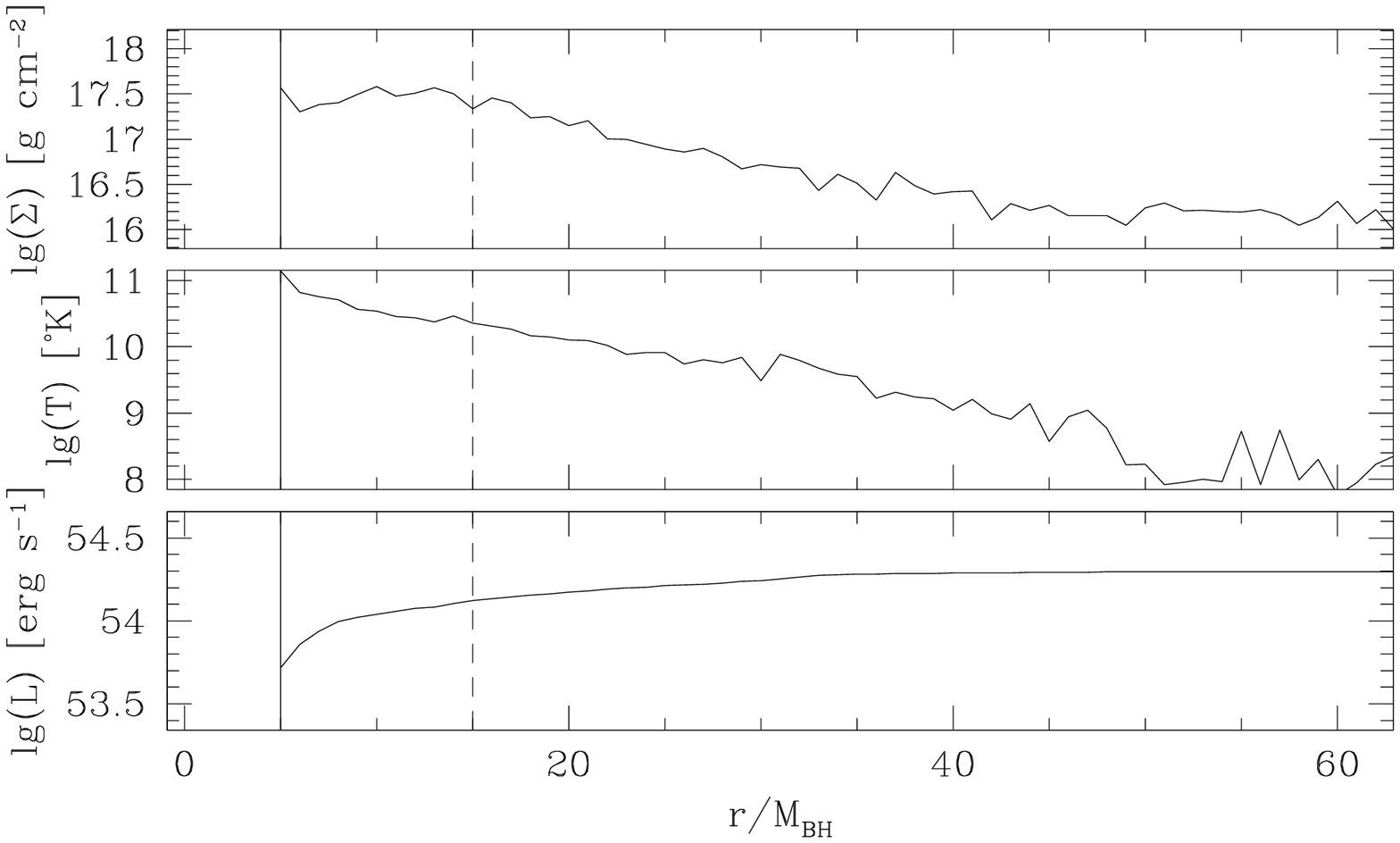}}
\caption{Surface density $\Sigma$, temperature $T$, and integrated
  neutrino luminosity $L$, as a function of
  cylindrical radius for the configuration shown in
  Fig.~\protect\ref{fig:xyz2}.  The solid vertical line denotes the
  ISCO, and the dashed line the transition radius between matter
  optically thick to neutrinos within and optically thin
  outside.\label{fig:rhoeps}}
\end{figure}

The specific internal energy in the inner part of the torus
corresponds to a temperature $T\approx 3-10 MeV\approx 2-7\times
10^{10} K$, shown in Fig.~\ref{fig:rhoeps}.  The surface density in this
region is $\Sigma\approx 2-3\times 10^{17} {\rm g~cm}^{-2}$.  Assuming
an opacity $\kappa=7\times 10^{-17}(T/10^{11} K)^2$, we
conclude that the disk is optically thick out to $r\sim 15M$.  In
the diffusion limit, the neutrino flux is $F_{\nu}\approx
7\sigma T^4/\kappa\Sigma$ where $\sigma$ is the Stefan-Boltzmann constant.
The total neutrino luminosity is $L_{\nu}\approx 2\pi r^2F_{\nu}\sim
10^{54}~{\rm erg~s}^{-1}$, which should be sufficient to generate the
annihilation luminosity required to power a SGRB.

Qualitatively, the hot torus described here is similar to that found
from collapsing HMNS remnants in NSNS mergers \cite{ShiUIUC}, 
but is physically larger.
While we do not follow the long-term evolution of the accretion torus
and surrounding material, we can estimate the fallback time for the
bound component, assuming geodesic orbits.  Approximately $0.03M_{\rm
NS}$ should return back toward the BH on timescales equal to or longer
than a second, which could in principle produce lower-energy bursts at
later times.  It is conceivable that this fallback accretion might
explain the secondary X-ray flares observed in SGRBs many seconds
after the initial burst (see \citep{Berger3} for a summary of the
observations), especially if self-gravity leads to the formation
of higher density clumps of material, but further simulations are
required to establish this identification.

\section{Conclusions}

Future observations should help to determine which progenitor
candidates are responsible for the observed short GRB population.
Gravitational wave measurements would provide important evidence if
detected in coincidence: A GRB resulting from hypermassive collapse
would occur noticeably delayed relative to the gravitational wave
signal from the inspiral and plunge phases of nearly equal-mass
objects, whereas one resulting from a BHNS binary would occur almost
immediately after the signal from a very unequal-mass merger.

To be able to predict which BHNS binaries would be SGRB progenitor
candidates, we need to determine the critical separation for tidal
disruption separating cases where the NS is swallowed whole from those
where a disk can form, which we have constrained to fall in the range
$3M_{\rm BH}<a_{R;crit}<6M_{\rm BH}$.  Of course, a proper study will
require us to drop the extreme mass ratio assumption and allow the BH
to propagate through the grid.  It is unclear that a very sharp
transition exists between cases where the NS disrupts just within or
just outside the ISCO, since both cases predict that the majority of
the NS mass accretes onto the BH while some is ejected outward.  We
are currently gearing up to study this process in fully general
relativistic simulations, relaxing both the assumptions of conformal
flatness and extreme mass ratios..

\section{Acknowledgments}
J.~A.~F. is supported by an NSF Astronomy and Astrophysics Postdoctoral
Fellowship under award AST-0401533.  T.~W.~B. gratefully acknowledges
support from the J.~S.~Guggenheim Memorial Foundation.  This work was
supported in part by NSF grants PHY-0205155 and PHY-0345151 and NASA
Grant NNG04GK54G to the University of Illinois, NSF Grant PHY-0456917
to Bowdoin College, and PHY-0245028 to Northwestern University.

\end{document}